\documentstyle[twocolumn,prl,aps,epsf]{revtex}
\begin{document}
\draft
\twocolumn[\hsize\textwidth\columnwidth\hsize\csname
@twocolumnfalse\endcsname
\title{Angular gaps in radial DLA:  two fractal dimensions
and non-transient deviations from linear self-similarity}
\author{Benoit B. Mandelbrot$^a$, Boaz Kol$^b$ and
Amnon Aharony$^b$}
\address{$^a$Department of Mathematics, Yale University, New Haven, CT 06520-8283}
\address{$^b$School of Physics and Astronomy, Raymond and Beverly Sackler
Faculty of Exact Sciences, \\ Tel Aviv University, Tel Aviv 69978,
Israel\\ }

\date{\today}
\maketitle
\begin{abstract}
When suitably rescaled, the distribution of the angular gaps
between branches of off-lattice radial DLA is shown to approach a
size-independent limit. The power-law expected from an asymptotic
fractal dimension $D=1.71$ arises only for very small angular
gaps, which occur only for clusters significantly larger than
$M=10^6$ particles. Intermediate size gaps exhibit an effective
dimension around $1.67$, even for $M \rightarrow \infty$. They
dominate the distribution for clusters with $M<10^6$. The largest
gap approaches a finite limit extremely slowly, with a correction
of order $M^{-0.17}$.
\end{abstract}
]

For years, a considerable effort has been devoted to the numerical
analysis of diffusion limited aggregation (DLA),\cite{Witten81} a
model of many natural fractal growth processes. Even though DLA
has no characteristic length scale, simulations exhibit nontrivial
and ambiguous characteristics.\cite{Mandelbrot95a,Mandelbrot95b}
In the plane, the fractal dimension $D=1.71$ is usually associated
with radial DLA,\cite{Tolman89} and $D=1.66$ is associated with
cylindrical DLA.\cite{Erzan95,Meakin86,Kol2000a,Kol2000b}

\begin{figure}
\centerline{ \epsfxsize 4.5cm \epsfbox{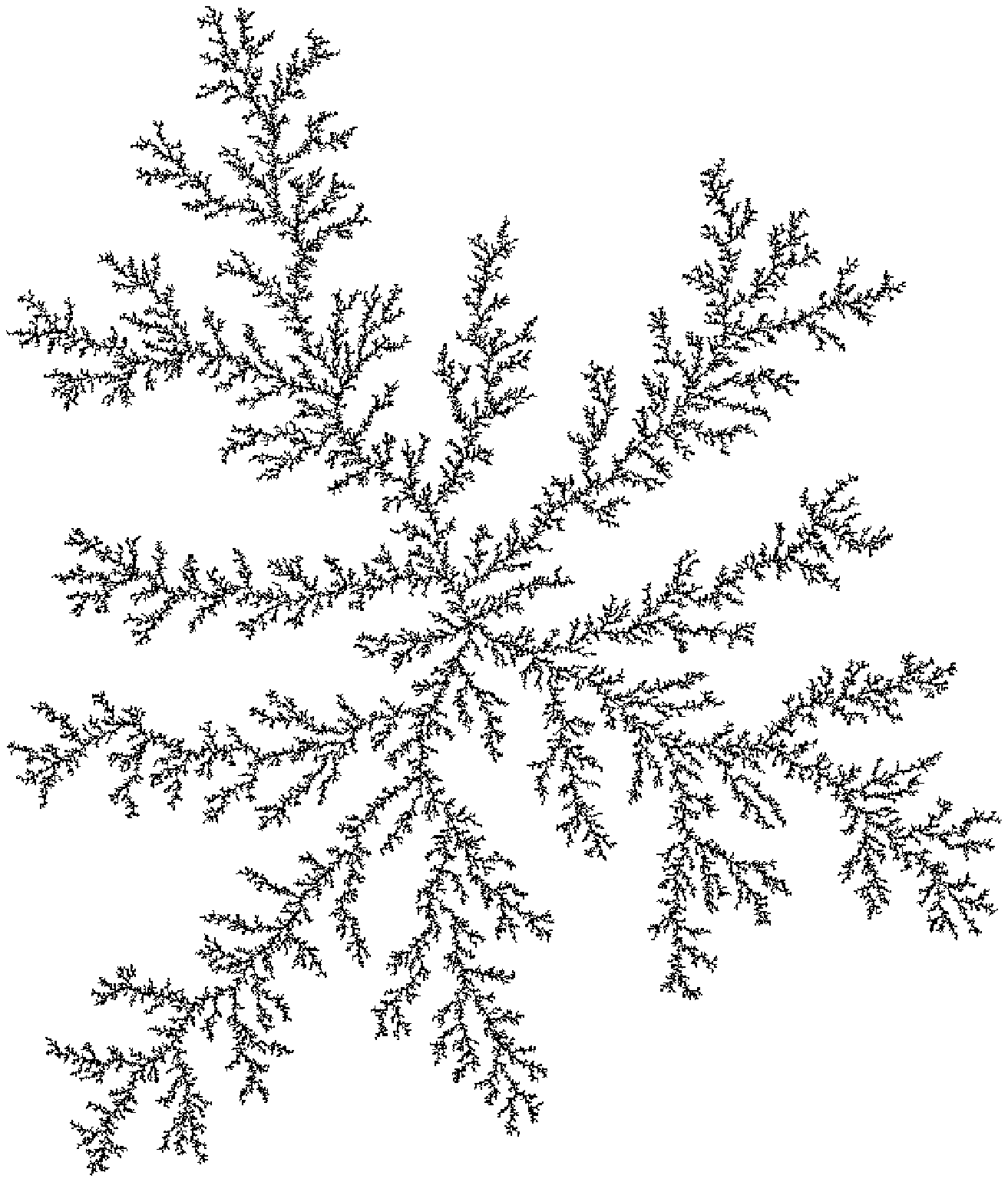}} \vspace{5mm}
\centerline{ \epsfxsize 4.5cm \epsfbox{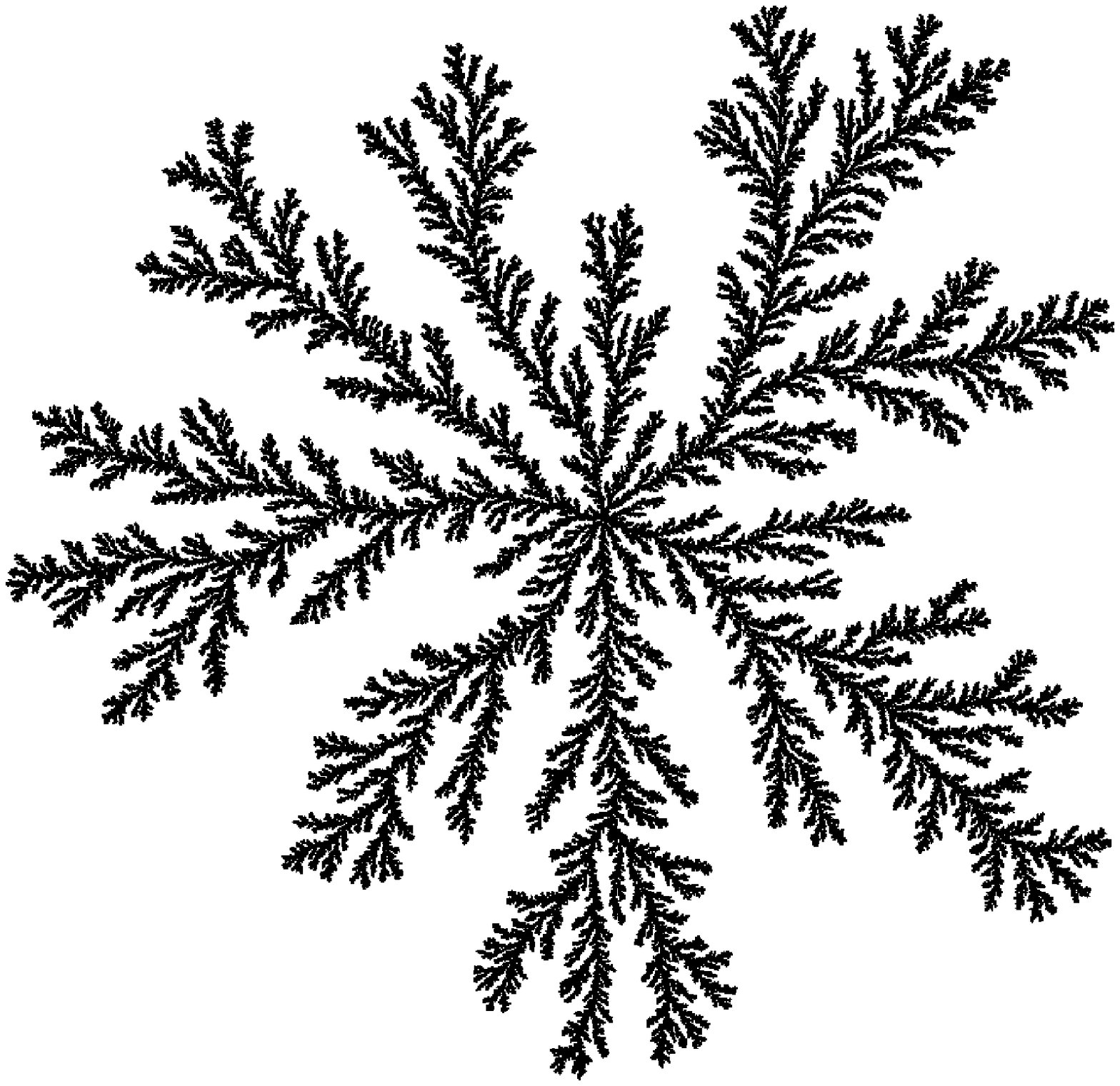}} \caption{Two
radial DLA clusters, with $M=10^5$ (top) and $M=10^8$ (bottom)
particles.} \label{DLAfigs}
\end{figure}

Although the mass of radial DLA obeys the relation $M \sim
R_g^{1.71}$ ($R_g$ is the radius of gyration) for all $R_g$'s, its
geometry varies with $R_g$.\cite{Mandelbrot92} Figure
\ref{DLAfigs} compares two DLA clusters: One recognizes about
${\cal N}=5$ main arms for $M=10^5$ particles,\cite{Arneodo92} but
${\cal N}=8$ or $9$ for $M=10^8$.\cite{Mandelbrot92} This makes
the $M=10^8$ cluster look more dense (like a filled circular
disc). \onlinecite{Mandelbrot95a,Mandelbrot95b}
%
However, different interpretations of ${\cal N}$ lead to
contradictory results. Lacking meaningful ways to analyze data,
there have been very few publications on numerical analysis of DLA
since 1995. Here we present a new systematic description that
incorporates the arms in a broad context: we study the statistics
of {\it all the angular gaps between branches}. While Ref.
\cite{Mandelbrot92} pioneered such gap analysis, it had
insufficient data and no data collapse. Application of the
``$\epsilon-$neighborhood" analysis related the increase in ${\cal
N}$ (and in the apparent density) to an increasing {\it
lacunarity}, which could remain consistent with $M \sim
R_g^{1.71}$. The two scenarios proposed there -- long transient
and infinite drift -- should be replaced by a subtle new one,
described here.

Our principal new finding is that, under appropriate scaling, and
apart from finite size tails, the distributions of gap sizes for
different $M$'s collapse onto a single function, which should also
characterize most of the gaps in the limit $M \rightarrow \infty$.
Intermediate size gaps exhibit an effective dimension around
$1.67$, and one needs very small gaps, hence very large clusters,
to observe the widely anticipated asymptotic behavior with the
unique dimension $D=1.71$. Interestingly, the two widely quoted
values of $D$ are {\it both} present in radial DLA.

The intermediate dimension  $D \sim 1.67$ is {\it not} a small
sample correction to asymptotic self-similarity. Indeed, the
number of gaps in this zone changes little with $M$, and they
remain observable for {\it all} $M$.
Therefore, DLA cannot be modeled by a unique fractal dimension,
that would correspond to a fractal that is invariant under {\it
linear} transformations. For many phenomena, like percolation,
random forms of such ``linear fractals" are good models and
non-random linear fractals provide illuminating ``cartoons". In
contrast, DLA (and probably other phenomena) require steps beyond
linear self-similarity. Our procedure suggests such a quantitative
step.

The data collapse of the gap distribution implies that typically,
the largest angular gap (which gives some measure of ${\cal N}$)
approaches a {\it finite} limit, which seems to be non-zero.
However, there is an amazingly slow correction, of order
$M^{-\gamma}$, where $\gamma = (2-D)/D=0.17$. This correction may
or may not be related to the corrections recently found using a
conformal map approach. \cite{Somfai99,Davidovitch2000b}

We performed large scale simulations of off-lattice radial DLA,
and focused on the gaps on circular cross-cuts. An earlier
analysis,\cite{Mandelbrot95b} based on $M \le 10^7$, yielded an
apparent fractal dimension $D=1.65 \pm 0.01$, which is not
consistent with the widely accepted $D=1.71$. Our software is as
described in Ref. \cite{Kaufman95}, but uses a single diffusing
particle each time, as in the definition of DLA.\cite{Witten81} We
generated $N_c(M)=1000$ clusters for each of the `masses'
$M=10^5,~10^{5.5},~10^6,~10^{6.5}$ and $10^7$, plus $N_c(M)=60$
additional clusters for $M=10^{7.5}$ and for $10^8$.

After growing several DLA clusters, we endeavored to keep to the
edge of the non-growing region. Trial and error
\cite{Mandelbrot95a,Mandelbrot95b} singled out particles whose
distance $R$ from the origin is within $0.75R_g-1<R<0.75R_g+1$ (in
units of the particle diameter).
The gap $\theta$ is the difference in the polar angles between
consecutive particle centers. To avoid ``gaps" between touching
particles, we keep only the values $\theta>1/0.75R_g$, larger than
one particle diameter. We then pool the gaps from all the
available clusters of size $M$ into a single list, which is sorted
from largest to smallest. Thus each gap is characterized by its
size and by an integer $r$ that represents its rank in the sorted
list.
Here we use the mean-rank,
$\rho \equiv r/N_c(M)$,
which becomes an $M-$independent continuous variable for $N_c
\rightarrow\infty$.

\begin{figure}
\centerline{ \epsfysize 5cm \epsfxsize 6cm \epsfbox{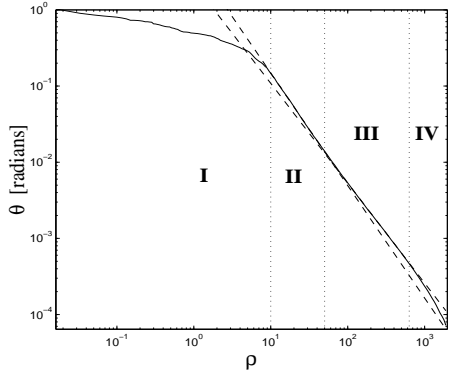}}
\vspace{1mm} \caption{The gap size $\theta$ vs. the mean-rank
$\rho$. The graph is based on $60$ clusters with $10^8$ particles.
The dotted vertical lines separate the four regions discussed in
the text. Dashed straight lines run through Regions II and III.}
\label{thetaRho8Fig}
\end{figure}

For an ideal linearly self-similar fractal of dimension $D$, one
expects a power law, $\theta \propto \rho^{-\alpha}$, where
$\alpha \equiv 1/D_c$ and $D_c\equiv D-1$ is the fractal dimension
of the intersection.\cite{Mandelbrot82} In contrast, Fig.
 \ref{thetaRho8Fig} shows that $\theta(\rho,M=10^8)$, based on
the available $N_c(10^8)=60$ clusters, exhibits four distinct
regions: Region I, with $\rho<10$, reflects the statistics of the
largest gaps. As in the Zipf-Mandelbrot
distribution,\cite{Mandelbrot82} the flattening of the curve in
this region is a token of an upper cutoff (though not necessarily
a sharp one) on the size of gaps.  Both Regions II and III appear
to be linear (see dashed lines in Fig. \ref{thetaRho8Fig}). In
Region II, with $10<\rho<50$, the apparent slope is $\alpha_{II}
\cong 1.5$, corresponding to $D_{II} \cong 1.67$.
In Region III, with $50<\rho<\rho_{\max}/3$, where
$\rho_{\max}(M=10^8)=1890$ is the maximal $\rho$ for which
$\theta$ is defined (i.e. the average total number of gaps), the
apparent slope is $\alpha_{III} \cong 1.33$, corresponding to
$D_{III} \cong 1.75$. Region IV, with
$\rho_{\max}/3<\rho<\rho_{\max}$, reflects finite-size
corrections: as $M \rightarrow \infty$ one has
$\rho_{\max}\rightarrow \infty$, and (as far as we can tell)
Region III extends to infinity, reflecting the asymptotic behavior
of DLA. Since $\rho_{\max}$ increases with $M$, it also follows
that for $M$ smaller than about $10^{6}$, Region III merges with
Region IV, and one would mistakenly identify $D$ from the higher
apparent slope of Region II. This clearly demonstrates why one
must have very large clusters to make statements about the
asymptotic behavior of DLA.

Visually, Region I defines the arms. As $M \rightarrow \infty$,
Region III collapses into the observed thickening `trunks'. Region
II determines the structure of the clearly separate intermediate
branches, which exist for all $M$.

\begin{figure}
\centerline{ \epsfysize 4cm \epsfxsize 6cm \epsfbox{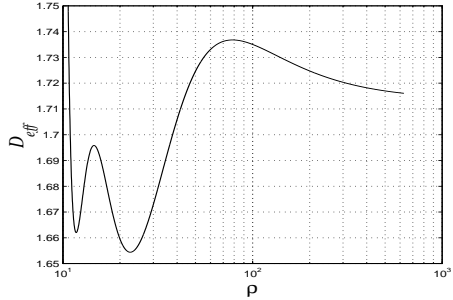}}
\vspace{1mm} \caption{Effective dimension $D_{\rm eff}(\rho)$.}
\label{M8LocalDimensionFig}
\end{figure}

A quantitative way to describe the crossover between Region II and
III would involve corrections to the asymptotic power law. Lacking
enough data, we represent the data we have by an ad-hoc parametric
power expansion,
$\theta(\rho)=\sum_{k=1}^{k_{\rm
max}}a_k\rho^{-k\alpha}=\sum_{k=1}^{k_{\rm max}}a_kx^k$,
where $x=\rho^{-\alpha}$. Using $\alpha=1.4$, this form ensures
the asymptotic dimension $D=1.71$ for large $\rho$. A least
squares fit with $k_{\rm max}=6$ to the data for
$10<\rho<\rho_{\rm max}/3$ at $M=10^8$ yields $a_k=3.702, -126.7,
1.566 \times 10^{4}, -8.045 \times 10^5, 1.901 \times 10^7, -1.694
\times 10^8$. The approximate effective dimension, $D_{\rm
eff}(\rho)\equiv1+1/\alpha_{\rm eff}$, where $\alpha_{\rm
eff}(\rho)\equiv -d(\log\theta)/
d(\log\rho)=\alpha\frac{x}{\theta}\frac{d\theta}{dx}$, is shown in
Fig. \ref{M8LocalDimensionFig}. Indeed, $D_{\rm eff}$ reaches a
minimum of about $1.66$ close to $\rho=20$ (in Region II), and a
maximum of about $1.74$ around $\rho=80$ (in the beginning of
Region III). It approaches the asymptotic $D=1.71$ at larger
$\rho$.

\begin{figure}

\centerline{ \epsfysize 5cm \epsfxsize 6cm \epsfbox{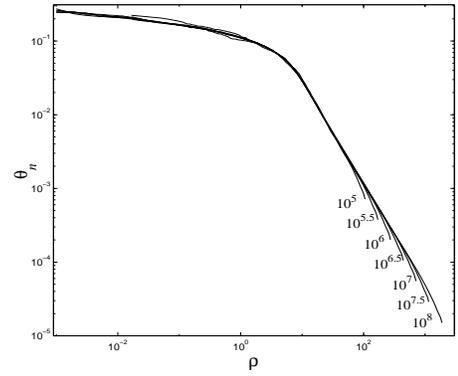}}
\vspace{1mm} \caption{Data collapse of $\theta_n(\rho,M)$. The
numbers indicate the cluster size $M$.} \label{GapDistributionFig}
\end{figure}

To demonstrate scaling and data collapse, we scale $\theta$ as
\begin{equation}
\theta(\rho,M)=S(M)\theta_n(\rho), \label{collapseEq}
\end{equation}
\begin{equation}
S(M)\equiv\int_0^G \theta(\rho,M)\, {\rm d}\rho. \label{thetan}
\end{equation}
$S(M)$ represents the largest gaps. $G$ should be large enough in
order to filter more of the statistical variance, but not too
large as to include finite-size corrections. The results are not
very sensitive to $G$ (see below). The choice $G=50$, at the
beginning of the asymptotic slope (Region III), gives good results
and facilitates some of the theoretical analysis presented later.
Figure \ref{GapDistributionFig} shows that $\theta_n(\rho,M)$ is
indeed independent of $M$, except for finite-size correction
deviations.

\begin{figure}
\centerline{
\epsfysize 6cm
\epsfbox{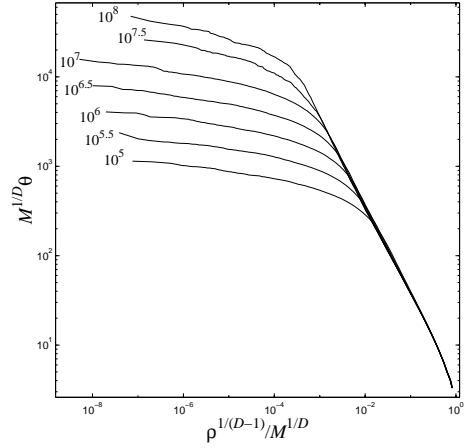}}
\vspace{1mm}
\caption{$M^{1/D}\theta$ vs. $\rho^{1/(D-1)}/M^{1/D}$ for various
values of $M$.} \label{finiteSizeCollapseFig}
\end{figure}

The curves in Fig. \ref{GapDistributionFig} end at $\rho_{\rm
max}(M)$. The figure clearly shows fixed distances between
consecutive end-points, indicating power-law dependence of both
$\rho_{\rm max}$ and of $\theta(\rho_{\rm max})$ on $M$. Since
$\rho_{\rm max}$ is the average number of gaps (or particles) on
the annulus of radius $R$, we expect $\rho_{\rm max}\propto
R^{D-1} \propto M^{(D-1)/D}$. The smallest gap is then expected to
scale as $1/R \propto 1/M^{1/D}$. We thus expect the small gaps to
scale as
\begin{equation}
\theta(\rho,M)=M^{-1/D}f(\rho^{1/(D-1)}/M^{1/D}).
\label{scalingRelation}
\end{equation}
Indeed, Fig. \ref{finiteSizeCollapseFig} shows $M^{1/D}\theta$
plotted vs. $\rho^{1/(D-1)}/M^{1/D}$ for various $M$'s. There is a
clear data collapse on the right-hand side, in agreement with Eq.
(\ref{scalingRelation}).

We next use the collapse of the normalized gaps in Fig.
\ref{GapDistributionFig} to infer the asymptotic distribution of
the original (non-normalized) gaps as a function of $M$. As
expected for $\rho<\rho_{\max}/3$, $\theta_n$ depends only on
$\rho$, and {\it not} on $M$. The $M$-dependence of the original
gaps, $\theta(\rho,M)= S(M)\theta_n(\rho)$, is thus determined by
the scale factor $S(M)$. Integrating over $\rho$ on both sides of
Eq. (\ref{collapseEq}), the left hand side becomes
$\int_0^{\rho_{\max}}\theta(\rho,M)=2\pi-X$, equal to the average
sum (over all cluster realizations) of all the gaps; $X$ is the
sum over the gaps that are smaller than one particle diameter.
We expect $X \propto \rho_{\max}/R_g$, and therefore
$X=B_1M^{-\gamma}$, with $\gamma\equiv (2-D)/D=0.17$. On the right
hand side, we divide the integral into three parts: $0<\rho<G$,
$G<\rho<\rho_{\rm max}/3$ and $\rho_{\rm max}/3<\rho<\rho_{\rm
max}$. Equation (\ref{thetan}) implies that $\int_0^G \theta_n
d\rho \equiv 1$. The second integral belongs to Region III, where
we write $\theta_n(\rho) \simeq A_1\rho^{-\alpha}$, and
$A_1=a_1/S(10^8)=0.77\pm 0.01$ is independent of $M$ (as seen from
Fig. \ref{GapDistributionFig}). This yields
$B_2[G^{-\beta}-(\rho_{\rm max}/3)^{-\beta}]$, with
$\beta=(2-D)/(D-1)$ and $B_2=A_1/\beta$. Since $\rho_{\rm max}
\propto M^{(D-1)/D}$, we conclude that the integral over the right
hand side is equal to $S(M)(A-B_3M^{-\gamma})$, where
$A=1+B_2G^{-\beta} =1.40\pm0.01$ and $B_3$ is a constant (which is
also affected by $\int_{\rho_{\rm max}/3}^{\rho_{\rm max}}\theta_n
d\rho \sim M^{-\gamma}$, due to Eq. (\ref{scalingRelation})). Our
conclusion is that for large $M$ one has
\begin{equation}
S(M)=\frac{2\pi-B_1M^{-\gamma}}{A-B_3M^{-\gamma}},
\label{SM}
\end{equation}
and hence
$\theta(\rho,M=\infty)=(2\pi/A)\theta_n(\rho)\cong
4.5\theta_n(\rho)$.

While the distribution of gaps does converge to a limit for
$M\to\infty$, this convergence is {\it amazingly} slow, being
governed by the small exponent $\gamma$: The correction term in
Eq. (\ref{SM}) reduces only by a factor of
$10^{3\times0.17}\simeq3.2$, between $M=10^5$ and $M=10^8$! This
means that great caution should be taken when extrapolating
various measurements in DLA, and that the
asymptotic limit
is attained only for extremely large $M$.

We next examine the statistics of the largest gaps, $\theta_{\rm
max}$. For a linearly self-similar fractal, one expects to find
the same distribution of the $\theta_{\rm max}$'s for different
cluster sizes. $\theta_{\rm max} \approx 2\pi/{\cal N}$ may
indicate that the number of main branches is around ${\cal N}$. It
has been shown that $\theta_{\rm max}$ decays slowly with
$M$,\cite{Mandelbrot95b} which means that ${\cal N}$ increases,
raising doubts on the self-similarity of DLA at intermediate
values of $M$.

If, instead of pooling, we had studied the gaps of a single
cluster, then we would have $\theta_{\rm max}=\theta(r=1)$.
However, the mean-rank $\rho=1$ has a different meaning:
$\theta(\rho=1,M)$ is the $N_c$'th largest gap out of all the gaps
taken from $N_c$ clusters. Since the graph of $\theta_n(\rho,M)$
is relatively flat near $\rho=1$, we still expect
$\theta(\rho=1,M)$ to represent the order of magnitude of the
average largest gap. Reading $\theta_n(\rho=1)=0.1$ from the
universal curve in Fig. \ref{GapDistributionFig}, we estimate
$\theta(\rho=1,M=\infty)=(2\pi/A)\theta_n(1) \cong 0.45$, in
radians, yielding an asymptotic average maximal gap around
$26^{\circ}$ and implying a finite number of arms.

\begin{figure}
\centerline{ \epsfysize 5cm \epsfxsize 6cm \epsfbox{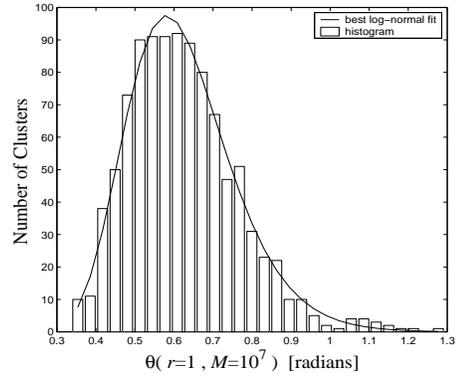}}
\vspace{1mm} \caption{Histogram of the maximal gap (in radians) in
each cluster, for the $1000$ available clusters with $M=10^7$.}
\label{Gap1HistogramFig}
\end{figure}

The histogram of the $\theta_{\rm max}$'s for different clusters
with $M=10^7$ is shown in Fig. \ref{Gap1HistogramFig}. It is quite
broad. A fit to a log-normal form (also shown) yields $\mu=\langle
\ln\theta_{\rm max}\rangle \cong-0.50$ and
$\sigma=\langle(\ln\theta_{\rm max}-\mu)^2\rangle\cong0.22$. We
find a very slow decrease of $\mu$ with $M$, approaching a finite
limiting value of the order of magnitude given above.

The broad distribution of $\theta_{\rm max}$ implies that under
pooling, the largest pooled gaps spread over a broad range of
$\rho$. Pooling convolutes the distributions of two
non-independent quantities: the gaps for each cluster and the
different clusters. For additional insight into these
distributions, we also considered the correlations between gaps
within each cluster. We find several interesting results: the
average correlation between the largest gap and the $r$'th gap for
$M=10^7$ turns out to decrease from 1 to 0 (around $r=5$), and to
approach a constant around $-0.5$ for $r>10$.
Negative correlations were expected, since the total sum of the
gaps must equal $2\pi$.

In collaboration of Eugene Vilensky of Yale, we also tested our
results' sensitivity to the special choice of $R \sim 0.75 R_g$.
An increase of $R$ beyond this value first leads to little change
(confirming \cite{Mandelbrot95a,Mandelbrot95b}), then to opposite
changes in different regions:  Very large gaps come to dominate in
Region I, moving the corresponding ``flat" portion of the curve to
higher values of $\theta$. The remaining gaps (other than the few
largest) add to a decreasing total angle, but with little change
in distribution.

In conclusion, we have identified robust structural features of
radial DLA, which act separately in characterizing the arms, the
intermediate and the fine branches. Although all fall within a
single scaling function, the non-trivial gap distribution function
reflects deviations from simple linear fractality. Since these
deviations characterize the radial geometry, it would be
interesting to check if they also appear in the (possibly simpler)
cylindrical case. This is particularly intriguing, since both
involve $D \approx 1.66$. Finally, since our data `only' extend up
to $M=10^8$, there is no guarantee that this is the end of the
story.\cite{DS} Future work should tell if larger scales bring
more surprise.

Henry Kaufman kindly adapted his DLA program to serve our needs.
This work has been supported in part by the German-Israeli Foundation
(GIF).

\end{document}